\documentclass[graybox,envcountsect]{svmult}

\usepackage{mathptmx}        % selects Times Roman as basic font
\usepackage{helvet}          % selects Helvetica as sans-serif font
\usepackage{courier}         % selects Courier as typewriter font
\usepackage{amssymb,amsmath}
\usepackage{cite}  
\usepackage[figuresright]{rotating}

\usepackage{makeidx}         % allows index generation
\usepackage{graphicx}        % standard LaTeX graphics tool
\usepackage{multicol}        % used for the two-column index
\usepackage[bottom]{footmisc}% places footnotes at page bottom

\newcommand{\nc}{\newcommand}   %%%%%% Just to make it easier to write new commands
\nc{\req}[1]{Eq.\,(\ref{#1})}    \nc{\reqp}[1]{Eq.\,(\ref{#1}) on page \pageref{#1}}     
\nc{\rf}[1]{Fig.~\ref{#1}}   \nc{\rfp}[1]{Fig.~\ref{#1} on page \pageref{#1}}     
\nc{\Th}{\ensuremath{T_\mathrm{H}\,}}
\nc{\pp}{\ensuremath{p\!p\ }}
\nc{\pA}{\ensuremath{p A\ }}
\nc{\hAA}{\ensuremath{AA\ }}
\def\beq{\begin{equation}}
\def\eeq{\end{equation}}

\begin{document}
%%%%%%%%%%%%%%%%%%%%%%%%%%%%%%%%%%%%%%%%%%%%%%%%%%%%%%%%%%%
\title*{Hagedorn's  Hadron Mass Spectrum and the Onset of Deconfinement$^*$}

\author{Marek Ga\'zdzicki and Mark I. Gorenstein} 
\institute{Marek: Goethe-University, Frankfurt, Germany; and  Jan Kochanowski University, Kielce, Poland\\
Mark: Bogolyubov Institute for Theoretical Physics, Kiev, Ukraine; and Frankfurt Institute for Advanced Studies,  Frankfurt, Germany\\
$^*$Chapter in: R. Hagedorn and J. Rafelski {\it Melting Hadrons, Boiling Quarks} (Springer 2005),\\ Chapter references related to other chapters in this book}  
\maketitle\label{MarekSec}
\vskip -1cm
\begin{abstract} 
{\hspace{3pt}A brief history of the observation of the onset of deconfinement - the beginning of the creation of quark gluon plasma in nucleus-nucleus collisions with increasing collision energy - is presented. It starts with the measurement of hadron mass spectrum and the Hagedorn's hypothesis of the limiting temperature of hadronic matter (the Hagedorn temperature). Then the conjecture that the Hagedorn temperature  is the phase transition temperature was formulated with the crucial Hagedorn participation. It was confirmed by the observation of the onset of deconfinement in lead-lead collisions at the CERN SPS energies.}
\end{abstract}

%%%%%%%%%%%%%%%%%%%%%%%%%%%%%%%%%%%%%%%%%%%%
\section{Hadron Mass Spectrum and the Hagedorn Temperature}

A history of multi-particle production  started with discoveries of hadrons, first in cosmic-ray experiments and soon after in experiments using beams of particles produced in accelerators. Naturally, the first hadrons, discovered in collisions of cosmic-ray particles, were the lightest ones, pion,  kaon and $\Lambda$. With the rapid advent of particle accelerators new particles were uncovered almost day-by-day. There are about 1000 hadronic states known so far. Their density in mass  $\rho(m)$  increases approximately exponentially as predicted by  the Hagedorn's Statistical Bootstrap Model~\cite{Hagedorn:1965stM} formulated in 1965:
\beq\label{rho-m}
\rho(m)~=~{\rm const}~m^{-a}\,\exp(b\,m)\;.
\eeq
In the case of point-like hadron states this leads to a single-particle partition function:
\beq\label{z}
Z(T,V)~=~\frac{V}{2\pi^2}\int_{m_\pi}^{\infty} 
dm\int_0^{\infty}k^2dk\, \exp\left(-\,\frac{\sqrt{k^2+m^2}}{T}\right)\,\rho(m)\;,
\eeq
where $V$ and $T$ are the system volume and temperature, respectively. 
The $m$-integral exists only for $ T < 1/b $. 
Thus, the hadron gas  temperature is limited from above. 
Its maximum temperature $\Th = 1/b$ (the so-called the Hagedorn temperature) 
was estimated by Rolf Hagedorn based on the 
1965 data to be $\Th\cong 160$~MeV. More recent estimates of the Hagedorn temperature(s) 
can be found in Ref.~\cite{Broniowski:2004yhM}, for further discussion see Chapters
 20 and 21. %\ref{3chap3C} and \ref{3chap3Cupdate}.

The first statistical model of multi--hadron production was proposed by Fermi~\cite{Fermi:1950jdM} in 1950. It  assumes that hadrons produced in high energy collisions are in equilibrium and that the energy density  of the created hadronic system increases with increasing collision energy. Soon after, Pomeranchuk~\cite{Pomeranchuk:1951eyM} pointed out that hadrons cannot decouple (freeze-out) at high energy densities. They will rather continue to interact while expanding until the matter density is low enough for interactions to be neglected. He estimated the freeze-out temperature to be close to pion mass, $\approx$150~MeV. Inspired by this idea Landau~\cite{Landau:1953gsM}, and his collaborators formulated a quantitative hydrodynamical model describing the expansion of  strongly interacting hadronic matter between the  Fermi's equilibrium high density stage (the early stage) and the  Pomeranchuk's low density decoupling stage (the freeze-out). The Fermi-Pomeranchuk-Landau picture serves as a base for modeling high energy nuclear collisions up to now~\cite{Florkowski_textbook}.

{\em The Hagedorn's conjecture concerning the limiting temperature was in  contradiction to the Fermi-Pomeranchuk-Landau model in which the temperature of hadronic matter created at the early stage of collisions increases monotonically with collision energy and it is unlimited. }

%%%%%%%%%%%%%%%%%%%%%%%%%%%%%%%%%%%%%%%%%%%%%%%%%%%%%
\section{Discovery of the Onset of Deconfinement}
The quark model of hadron classification proposed by Gell-Mann and  Zweig in 1964 
starts a 15 years-long period in which sub-hadronic particles, 
quarks and gluons, were discovered and a theory of their interactions, 
quantum chromodynamics (QCD) was established. 
In parallel, conjectures were formulated concerning the existence 
and properties  of matter consisting of sub-hadronic particles, 
soon called the QGP and studied in detail within the QCD~\cite{Shuryak:1980tpM}.

Ivanenko, Kurdgelaidze~\cite{Ivanenko:1965dgM}, Itoh~\cite{Itoh:1970uwM} and
Collins, Perry~\cite{Collins:1974kyM}
suggested that quasi-free quarks 
may exist in the centre of neutron stars. Many physicists started to speculate 
that the QGP can be formed in nucleus--nucleus collisions at high energies 
and thus it may be discovered in laboratory experiments. 
Questions concerning QGP properties and properties of its transition to 
matter consisting of hadrons were considered since the late 70s.

Cabibbo, Parisi~\cite{Cabibbo:1975igM} pointed out that the exponentially
increasing mass spectrum proposed by Hagedorn may be connected to the existance
of the phase in which quarks are not confined.
Then Hagedorn and Rafelski~\cite{Hagedorn:1980cvM}, see Chapter 23, %~\ref{3chap4},  
Gorenstein, Petrov, and Zinovjev~\cite{GPZ} suggested that the Hagedorn massive 
states are not the point-like objects but the quark-gluon bags. 
These picture leads to the interpretation of  the upper limit of the hadron gas temperature, 
the  Hagedorn temperature, as the transition temperature from the hadron gas to a quark gluon plasma.
Namely,  at $ T > \Th $ the temperature refers to an interior of the quark-gluon bag, i.e., to the QGP. 

In the mid-90s the Statistical Model of the Early Stage (SMES) was formulated~\cite{Gazdzicki:1998vdM} as an extension of the Fermi's statistical model of hadron production. It assumes a statistical production of confined matter at low collision energies (energy densities) and a statistical QGP creation at high collision energies (energy densities). The model predicts a rapid change of the collision energy dependence of hadron production properties, that are sensitive to QGP, as a signal of a transition to QGP (the onset of deconfinement) in nucleus--nucleus collisions. The onset energy was estimated to be located in the CERN SPS energy range.

{\em Clearly, the QGP hypothesis and the SMES model removed the contradiction between the Fermi's and Hagedorn's statistical approaches. Namely, the early stage temperature of strongly interacting matter is unlimited and increases monotonically with collisions energy, whereas there is a maximum temperature of the hadron gas, $\Th \approx 160$~MeV, 
above which strongly interacting matter is in the  QGP phase.}

Rich data from experiments at the CERN SPS and LHC as well as at the BNL AGS and RHIC clearly indicate that a system of strongly interacting particles created in heavy collisions at high energies is close to, at least local, equilibrium. At freeze-out the system occupies a volume which is much larger than a volume of an  individual hadron. Thus, one concludes that  strongly interacting matter is created in heavy ion collisions~\cite{Florkowski_textbook}.

%%%%%%%%%%%%%%%%%%%%%%%%%
\begin{figure}[tbh]
\centering
\includegraphics[width=0.49\columnwidth]{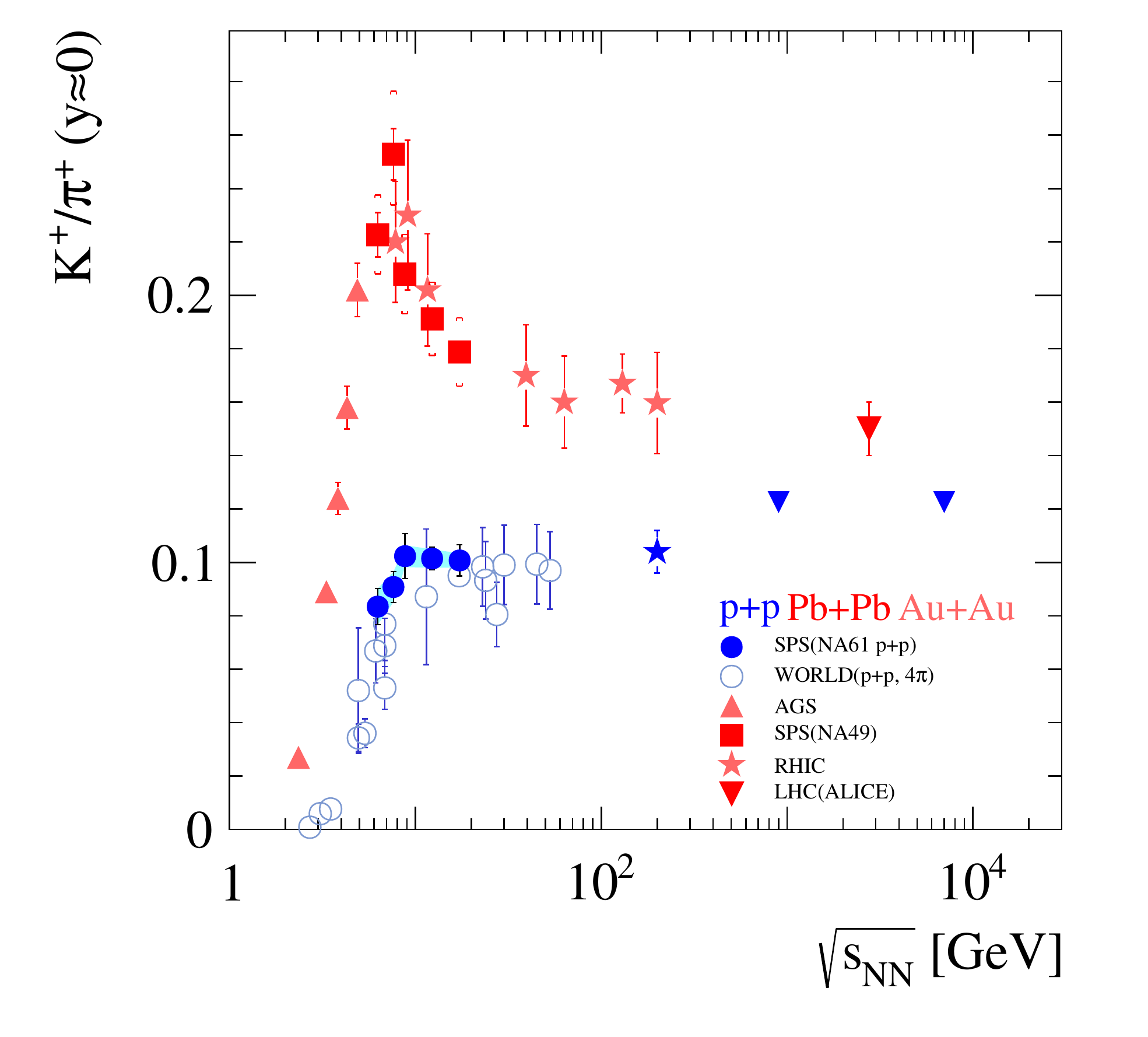}
\includegraphics[width=0.49\columnwidth]{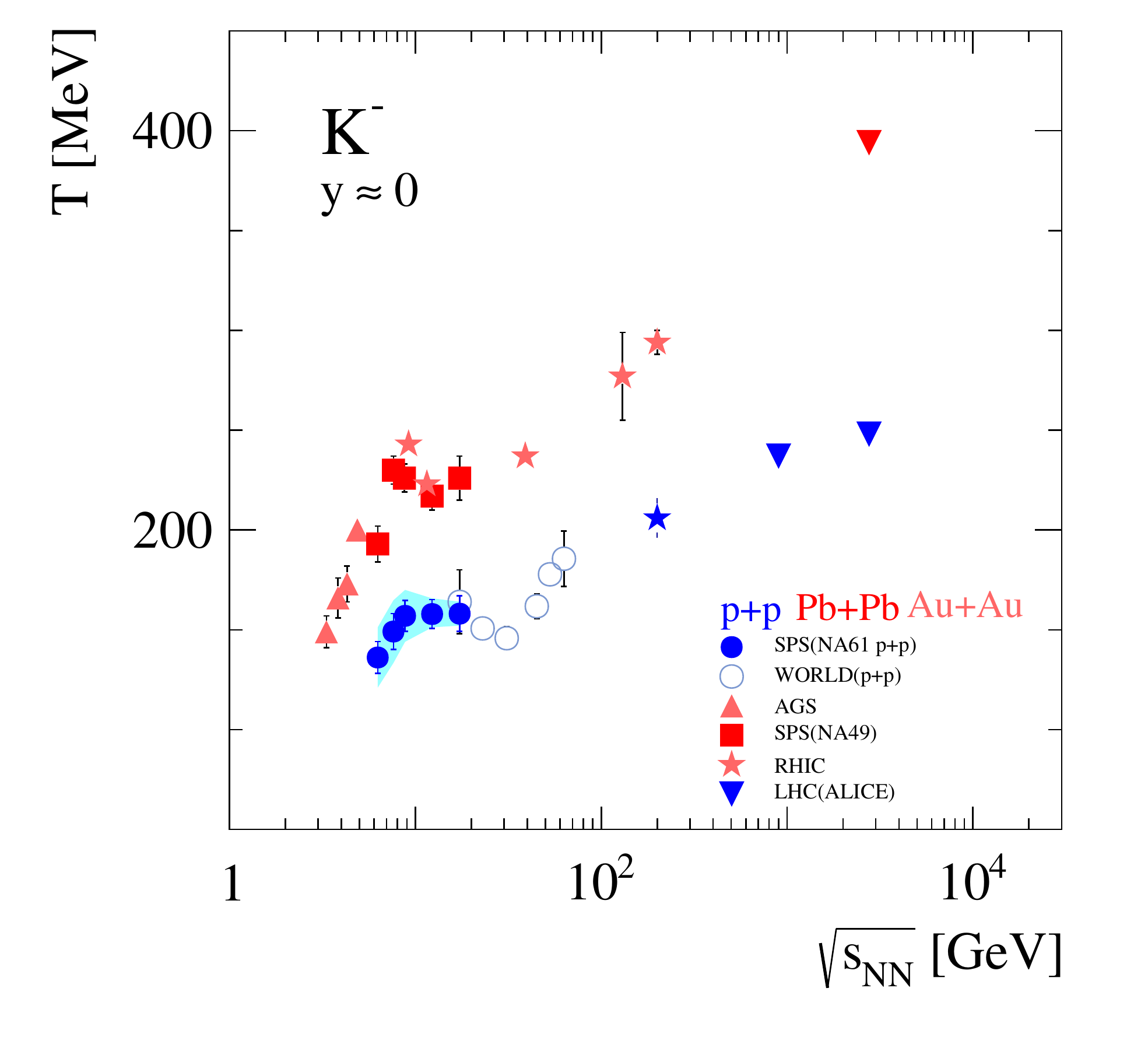}
  \caption{\label{fig:onset}
Recent results on the observation of the phase transition
in central Pb+Pb (Au+Au) collisions~\cite{sr2014}.
The horn (left) and step (right) structures
in energy dependence of the  K$^+/\pi^+$ ratio and the inverse slope
parameter of  K$^-$ $m_\bot$ spectra signal the onset of deconfinement
located at the low CERN SPS energies.
}
\end{figure}
%%%%%%%%%%%%%%%%%%%%%%%%%

The phase transition of strongly interacting matter to the QGP was discovered within the energy scan program of the NA49 Collaboration at the CERN SPS~\cite{Afanasiev:2002mxM,Alt:2007feM}. The program was motivated by the predictions of the SMES model. The discovery was based on the observation that several basic hadron production properties measured in heavy ion collisions rapidly change their dependence on collisions energy in a common energy domain~\cite{Gazdzicki:2010ivM}, see Fig.~\ref{fig:onset}.

\begin{figure}[th]
\vskip -0.5cm
\centering 
\includegraphics[width=0.95\columnwidth]{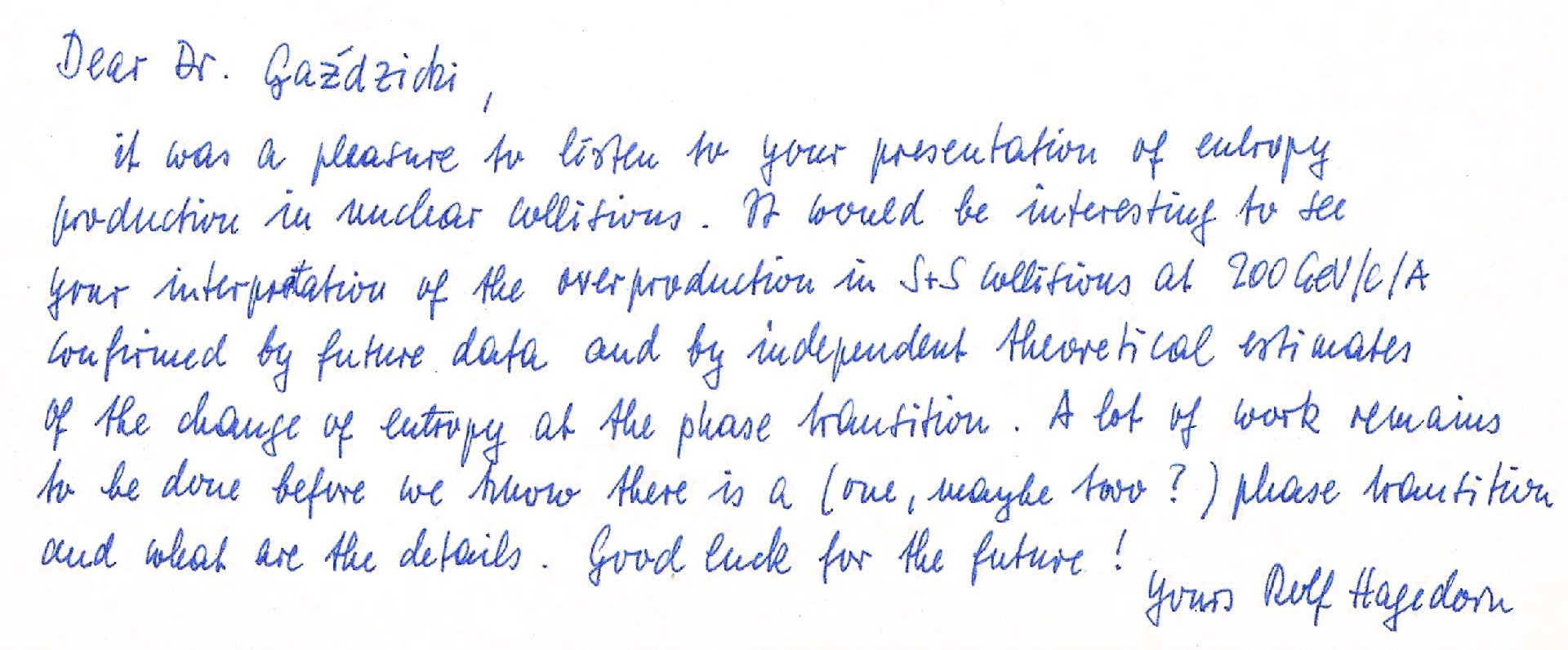}
  \caption{\label{fig:letter}
The letter of Rolf Hagedorn to Marek commenting the first talk on
the onset of deconfinement in nucleus-nucleus collisions at the
low CERN SPS energies~\cite{Gazdzicki:1994udM} presented in June 1994 at 
the Divonne workshop dedicated to  Rolf Hagedorn on occasion of his 75th birthday.
}
\end{figure}

The first ideas which resulted in formulation of the SMES model were presented by one of us~\cite{Gazdzicki:1994udM} at the Workshop on {\it Hot hadronic matter: Theory and experiment}, which took place in Divonne, France in June 1994. The workshop was dedicated to 75th birthday of Rolf Hagedorn. Hagedorn's letter on the presentation is reprinted in Fig.~\ref{fig:letter} in lieu of summary.
%%%%%%%%%%%%%%%%%%%%%%%%%
%%%%%%%%%%%%%%%%%%%%%%%%%
%%%%%%%%%%%%%%%%%%%%%%%%%
\begin{figure}%[th]
\centering 
\includegraphics[width=0.90\columnwidth]{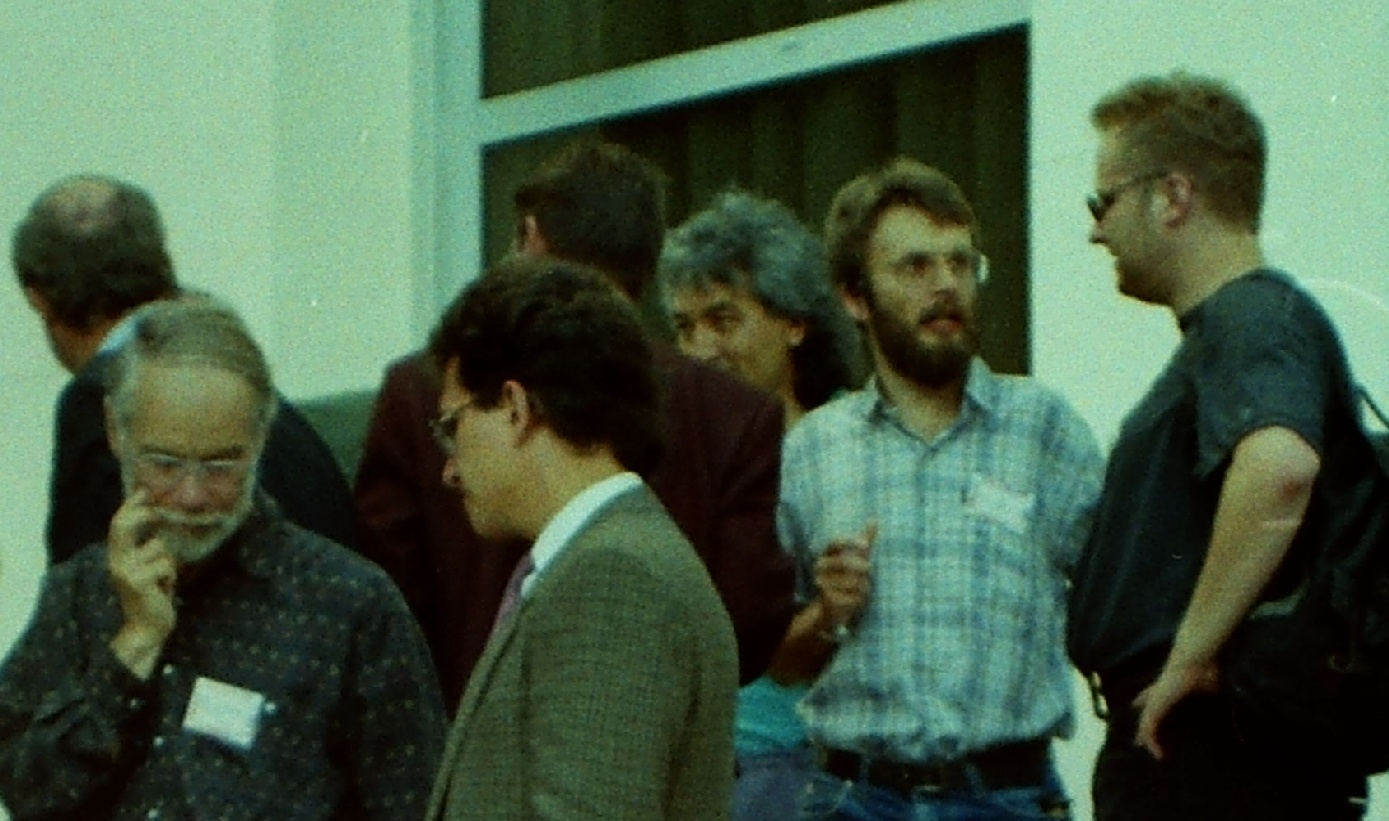}
  \caption{\label{fig:MarekDivonne}
Marek (facing to right off center) at Hagedorn Divonne Fest, June 30, 1994.
}
\end{figure}
%%%%%%%%%%%%%%%%%%%%%%%%%

{\acknowledgement{
This work was supported by
the National Science Centre of Poland (grant
UMO-2012/04/M/ST2/00816),
the German Research Foundation (grant GA 1480\slash 2-2) and
the
Program of Fundamental Research of the Department of Physics and
Astronomy of NAS, Ukraine.\\[-0.9cm]}}

%%%%%%%%%%%%%%%%%%%%%%%%%%%%%%%%%%%%%%%%%%%%%%%%%%%%%%%%%%%%%%%%%%%%%%%%%%%%%%%%%%%%%%%%%%%%%%%%%%%%

\end{document}